\documentclass[twocolumn,showpacs,preprintnumbers,amsmath,amssymb]{revtex4}
\usepackage{graphicx}% Include figure files
\usepackage{dcolumn}% Align table columns on decimal point
\usepackage{bm}% bold math

\begin{document}

\preprint{}

\title{Entanglement
Split:Comment on ``Quantum secret sharing based on reusable
Greenberger-Horne-Zeilinger states as secure carriers''
[Phys. Rev. A 67, 044302 (2003)]}% Force line breaks with \\

\author{Jian-Zhong Du$^{1,2}$, Su-Juan Qin$^{1}$, Qiao-Yan Wen$^{1}$, and Fu-Chen Zhu$^{3}$\\
        $^1$School of Science, Beijing University of Posts and Telecommunications, Beijing, 100876, China \\
        $^2$State Key Laboratory of Integrated Services Network, Xidian University, Xi'an, 710071, China \\
        $^3$National Laboratory for Modern Communications, P.O.Box 810, Chengdu, 610041, China \\ Email: ddddjjjjzzzz@tom.com}

\begin{abstract}
In a recent paper [S. Bagherinezhad and V. Karimipour, Phys. Rev.
A 67, 044302 (2003)], a quantum secret sharing protocol based on
reusable GHZ states was proposed. However, in this comment, it is
shown that this protocol is insecure because a cheater can gain
all the secret bits before sharing, while introducing one data bit
error at most in the whole communication, which makes the cheater
avoid the detection by the communication parities.
\end{abstract}


\maketitle

In a recent paper \cite {BK}, Bagherinezhad and Karimipour proposed
a quantum secret sharing protocol based on reusable GHZ states as
secure carriers (BK protocol). The security against both
intercept-resend strategy and entangle-ancilla strategy was proved.
Gao, Guo, Wen and Zhu showed that Eve can obtain the data bits in
the odd rounds without being detected by the communication parties
\cite {GGWZ}. To avoid this attack, the legitimate parties need to
interspace by stray random bits in the odd rounds \cite {K}.
However, we will show that a cheater, called Bob, can gain all the
secret bits before sharing at the cost of one data bit error at
most. Bob employs a special attack strategy called entanglement
split in one even round, and intercept-resends the
 data bits with the split entanglement in the sequel.

 Let us give a
brief description of the BK protocol \cite {BK}. In the odd round,
Alice, Bob and Charlie share the carriers
$|G\rangle_{abc}=(1/{\sqrt {2}})(|000\rangle+|111\rangle)_{abc}$.
Alice entangles the state $|qq\rangle_{12}$ that denotes the data
bit \emph{q} to $|G\rangle_{abc}$ by performing two CNOT gates
$C_{a1}C_{a2}$ to produce the state $|\Phi^{odd}\rangle=(1/{\sqrt
{2}})(|000\rangle_{abc}|qq\rangle_{12}+|111\rangle_{abc}|1+q,1+q\rangle_{12})$(CNOT
is specified by two subscripts, the first one is the control bit,
the second is the target bit.). In the even round, three parties
share the carriers
$|E\rangle_{abc}=(1/2)(|000\rangle+|110\rangle+|101\rangle+|011\rangle)_{abc}$.
Alice entangles the state $|\overline{q}\rangle_{12}=(1/{\sqrt
{2}})(|0,q\rangle+|1,1+q\rangle)_{12}$ that denotes the data bit
\emph{q} to $|E\rangle_{abc}$ by performing one single CNOT gate
$C_{a1}$ to produce the state $|\Psi^{even}\rangle=(1/{\sqrt
{2}})(|0\rangle_{a}|\overline{0}\rangle_{bc}|\overline{q}\rangle_{12}
+|1\rangle_{a}|\overline{1}\rangle_{bc}|\overline{1+q}\rangle_{12})$.
Then Alice transmits qubits \emph{1,2} to Bob and Charlie
respectively. At the destination, Bob performs one single CNOT
gate $C_{b1}$ and measures the qubit \emph{1} to obtain the
sending bit, and Charlie performs CNOT gate $C_{c2}$ and measures
the qubit \emph{2}. At the end of every round, Alice, Bob and
Charlie act the local operations of Hadamard gates on their
carries respectively to transform $|G\rangle$ and $|E\rangle$ into
each other, namely the appropriate one for the next round. Eve's
presence can be detected by publicly comparing a subsequence of
bits sent by Alice with those received by Bob and Charlie  after
the data bits are measured.

For convenience, we use the same notations as in Ref.\cite {BK}.

In the beginning, corresponding to the data bits  $q_1, q_2, q_3,
..., q_n$ that Alice wants to distribute to Bob and Charlie, Bob
prepares two variable strings  $d_1, d_2, d_3, ..., d_n$ and $e_1,
e_2, e_3, ..., e_n$, which are used to record the eavesdropping bits
and  the announcing bits respectively. For convenience, let the
cheat start the second round.

The whole cheat strategy consists of splitting entanglement,
maintaining the split entanglement, resending bit, and intercepting
bit.

1.The operation of splitting entanglement

The entanglement split occurs in the second round. Bob intercepts
the sending qubit \emph{2}, performs a unitary operation \emph{U} on
particles \emph{b, 1} and \emph{2}, and discards particle \emph{2},
 where $U|000\rangle=|000\rangle$,
$U|001\rangle=|110\rangle$, $U|010\rangle=|111\rangle$,
$U|011\rangle=|001\rangle$, $U|100\rangle=|100\rangle$,
$U|101\rangle=|010\rangle$, $U|110\rangle=|011\rangle$, and
$U|111\rangle=|101\rangle$. The analysis is as following.

Alice encodes her one bit into qubits \emph{1,2} to produce the
state
\begin{eqnarray}
|\Psi^0_{abc12}\rangle=\frac 1 {2\sqrt
{2}}(|00000\rangle+|00011\rangle+|01100\rangle+|01111\rangle\nonumber\\+|10101\rangle
+|10110\rangle+|11001\rangle+|11010\rangle) \ (q_2=0),\nonumber\\
or \ |\Psi^1_{abc12}\rangle=\frac 1 {2\sqrt
{2}}(|00001\rangle+|00010\rangle+|01101\rangle+|01110\rangle\nonumber\\+|10100\rangle
+|10111\rangle+|11000\rangle+|11011\rangle) \ (q_2=1).
\end{eqnarray}

Here we use superscripts 0 and 1 to denote the states corresponding
to $q_2=0$ and $q_2=1$, respectively. This notation also applies to
the following equations and we will, for simplicity, suppress the
word ``or'' later.

After Bob performing the unitary operation \emph{U} on particles
\emph{b, 1} and
 \emph{2}, the state will be converted into
\begin{eqnarray}
|\Theta^0_{abc12}\rangle=\frac 1 {2\sqrt
{2}}(|00000\rangle+|00001\rangle+|01100\rangle+|01101\rangle\nonumber\\+|11110\rangle
+|11111\rangle+|10010\rangle+|10011\rangle),\nonumber
\end{eqnarray}
\begin{eqnarray}
|\Theta^1_{abc12}\rangle=\frac 1 {2\sqrt
{2}}(|01010\rangle+|01011\rangle+|00110\rangle+|00111\rangle\nonumber\\+|10100\rangle
+|10101\rangle+|11000\rangle+|11001\rangle),
\end{eqnarray}
namely
\begin{eqnarray}
|\Theta^0_{a1bc2}\rangle=(1/{\sqrt{2}})(|00\rangle+|11\rangle)_{a1}\otimes
(1/{\sqrt{2}})(|00\rangle+|11\rangle)_{bc}\nonumber\\\otimes (1/{\sqrt{2}})(|0\rangle+|1\rangle)_2\nonumber,\\
|\Theta^1_{a1bc2}\rangle=(1/{\sqrt{2}})(|01\rangle+|10\rangle)_{a1}\otimes
(1/{\sqrt{2}})(|01\rangle+|10\rangle)_{bc}\nonumber\\\otimes
(1/{\sqrt{2}})(|0\rangle+|1\rangle)_2.
\end{eqnarray}

The entanglement split occurs. The pair of qubit \emph{a} and qubit
\emph{1} is the EPR pair between Alice and Bob. The pair of qubit
\emph{b} and
 qubit \emph{c} is the EPR pair between Bob and Charlie.
The carriers become two EPR pairs instead of the initial GHZ
state.

 This qubit \emph{1} in the second round is one part of the carries
 in the sequel, but the qubit \emph{1} in another round will be
 measured by Bob. For distinction, we regards this qubit \emph{1} as
 qubit $\overline
{b}$, which is always in Bob's site in the sequel.

2.The operation of maintaining the split entanglement

In every round, after an encoding operation and the corresponding
decoding operation are finished by Alice and Bob, or by Bob and
Charlie, the carries of two EPR states will be restored. The
analysis is in the operation of resending bit  and the operation of
intercepting bit.

 According to the BK protocol in Ref.\cite {BK}, Alice
and Charlie perform two Hardamard gates on their qubits \emph{a} and
\emph{c} respectively at the end of every round. To maintain the
split entanglement, Bob performs the same actions on qubits \emph{b}
and $\overline {b}$ respectively. We write the evolvement by
\begin{eqnarray}
H_a\otimes H_{\overline{b}}\otimes H_b\otimes H_c\frac 1 {\sqrt
{2}}(|00\rangle+|11\rangle)_{a\overline{b}}\otimes \frac 1 {\sqrt
{2}}(|00\rangle+|11\rangle)_{bc}\nonumber\\
=\frac 1 {\sqrt {2}}(|00\rangle+|11\rangle)_{a\overline{b}}\otimes
\frac 1 {\sqrt {2}}(|00\rangle+|11\rangle)_{bc},\nonumber
\\
H_a\otimes H_{\overline{b}}\otimes H_b\otimes H_c\frac 1 {\sqrt
{2}}(|01\rangle+|10\rangle)_{a\overline{b}}\otimes \frac 1 {\sqrt
{2}}(|01\rangle+|10\rangle)_{bc}\nonumber\\
=\frac 1  {\sqrt {2}}(|00\rangle-|11\rangle)_{a\overline{b}}\otimes
\frac 1 {\sqrt {2}}(|00\rangle-|11\rangle)_{bc},\nonumber\\
 H_a\otimes
H_{\overline{b}}\otimes H_b\otimes H_c\frac 1 {\sqrt
{2}}(|00\rangle-|11\rangle)_{a\overline{b}}\otimes \frac 1 {\sqrt
{2}}(|00\rangle-|11\rangle)_{bc}\nonumber\\
=\frac 1 {\sqrt {2}}(|01\rangle+|10\rangle)_{a\overline{b}}\otimes
\frac 1 {\sqrt {2}}(|01\rangle+|10\rangle)_{bc}.
\end{eqnarray}

As a result, if $q_2=0$ the carriers will be $(1/{\sqrt
{2}})(|00\rangle+|11\rangle)_{a\overline {b}}\otimes(1/{\sqrt
{2}})(|00\rangle+|11\rangle)_{bc}$ in every round, whereas if
$q_2=1$ the carriers will be $(1/{\sqrt
{2}})(|00\rangle-|11\rangle)_{a\overline {b}}\otimes (1/{\sqrt
{2}})(|00\rangle-|11\rangle)_{bc}$ in every odd round but $(1/{\sqrt
{2}})(|01\rangle+|10\rangle)_{a\overline {b}}\otimes (1/{\sqrt
{2}})(|01\rangle+|10\rangle)_{bc}$  in every even round.

3.The operation of resending bit

From the analysis above, In every round, Bob and Charlie share one
of three kinds of EPR carries:
$(1/{\sqrt{2}})(|00\rangle+|11\rangle)_{bc}$,
$(1/{\sqrt{2}})(|00\rangle-|11\rangle)_{bc}$, and
$(1/{\sqrt{2}})(|01\rangle+|10\rangle)_{bc}$.

 To resend a bit $\psi$, Bob prepares  one
counterfeit particle in the state $|\psi\rangle$ as qubit \emph{2}.
Bob entangles the state $|\psi\rangle_{2}$
 to the carriers by
performing one CNOT gate $C_{b2}$  on
\begin{eqnarray}
(1/{\sqrt{2}})(|00\rangle+|11\rangle)_{bc}|\psi\rangle_2,\nonumber\\
(1/{\sqrt{2}})(|00\rangle-|11\rangle)_{bc}|\psi\rangle_2,\nonumber\\
(1/{\sqrt{2}})(|01\rangle+|10\rangle)_{bc}|\psi\rangle_2
\end{eqnarray}
to produce the state
\begin{eqnarray}
(1/{\sqrt{2}})(|0,0,\psi\rangle+|1,1,1+\psi\rangle)_{bc2},\nonumber\\
(1/{\sqrt{2}})(|0,0,\psi\rangle-|1,1,1+\psi\rangle)_{bc2},\nonumber\\
(1/{\sqrt{2}})(|0,1,\psi\rangle+|1,0,1+\psi\rangle)_{bc2}.
\end{eqnarray}
Then Bob transmits the counterfeit qubit to Charlie.

At the destination, according to the BK protocol in Ref.\cite {BK},
Charlie acts on this state by the operator $C_{c2}$ to produce the
state
\begin{eqnarray}
(1/{\sqrt{2}})(|00\rangle+|11\rangle)_{bc}|\psi\rangle_2,\nonumber\\
(1/{\sqrt{2}})(|00\rangle-|11\rangle)_{bc}|\psi\rangle_2,\nonumber\\
(1/{\sqrt{2}})(|01\rangle+|10\rangle)_{bc}|1+\psi\rangle_2.
\end{eqnarray}

Charlie measures the counterfeit particle to extract the
transmission bit. The bit sent by Bob and that received by Charlie
are same when Bob and Charlie share the carriers $(1/{\sqrt
{2}})(|00\rangle \pm|11 \rangle)_{bc}$, but inverse when they share
$(1/{\sqrt {2}})(|01\rangle+|10\rangle)_{bc}$.

In the second round, after the operation of splitting entanglement,
Bob discards qubit \emph{2}, then records $e_2=0$, $d_2=0$ and
$\psi=0$. Bob resends the bit $\psi$ to Charlie. At the destination,
Charlie receives the bit 0 when $q_2=0$, but the bit 1 when $q_2=1$.

4.The operation of intercepting bit and the analysis of the cheat
strategy

Firstly, consider the intercept-resend in every odd round (round
$2m+1$, $3\leq 2m+1\leq n$).

According to the BK protocol in Ref.\cite {BK}, after Alice's
operations $C_{a1}C_{a2}$, the whole state will be converted into
\begin{eqnarray}
(1/{\sqrt{2}})(|00\rangle_{a\overline{b}}|q,q\rangle_{1,2} +
|11\rangle_{a\overline{b}}|1+q,1+q\rangle_{1,2})\nonumber\\
\otimes (1/{\sqrt{2}})(|00\rangle + |11\rangle)_{bc}\ (q_2=0),\nonumber\\
(1/{\sqrt{2}})(|00\rangle_{a\overline{b}}|q,q\rangle_{1,2} -
|11\rangle_{a\overline{b}}|1+q,1+q\rangle_{1,2})\nonumber\\
\otimes (1/{\sqrt{2}})(|00\rangle - |11\rangle)_{bc}\ (q_2=1).
\end{eqnarray}

After receiving qubit \emph{1} and intercepting qubit \emph{2}, Bob
performs CNOT gates $C_{\overline{b} 1}C_{\overline{b} 2}$ on qubits
$\overline{b}$,\emph{1} and \emph{2} to produce the state
\begin{eqnarray}
\frac 1 {\sqrt {2}}(|00\rangle +|11\rangle)_{a\overline{b}}\otimes
|q,q\rangle_{1,2} \otimes \frac 1 {\sqrt {2}}(|00\rangle +
|11\rangle)_{bc}\ (q_2=0),\nonumber
\\
\frac 1 {\sqrt {2}}(|00\rangle -|11\rangle)_{a\overline{b}}\otimes
|q,q\rangle_{1,2} \otimes \frac 1 {\sqrt {2}}(|00\rangle -
|11\rangle)_{bc}\ (q_2=1).
\end{eqnarray}
Then he measures qubits \emph{1,2} in the basis \{$|00\rangle$,
$|01\rangle$, $|10\rangle$, $|11\rangle$\}. He gains the measurement
outcome $|q,q\rangle_{12}$ that denotes bit q. The bit sent by Alice
and that received by Bob are always same.

Bob records $d_{2m+1}=q$, $e_{2m+1}=q$, and $\psi=q$. By the
 operation of resending bit, Bob sends bit $\psi$ to Charlie with the help of the
carriers $(1/{\sqrt {2}})(|00\rangle+|11\rangle)_{bc}$ or $(1/{\sqrt
{2}})(|00\rangle-|11\rangle)_{bc}$.

From the above analysis, we can see that because Alice, Bob, and
Charlie share the carriers $(1/{\sqrt
{2}})(|00\rangle+|11\rangle)_{a\overline {b}}\otimes (1/{\sqrt
{2}})(|00\rangle+|11\rangle)_{bc}$ or the carriers $(1/{\sqrt
{2}})(|00\rangle-|11\rangle)_{a\overline {b}}\otimes (1/{\sqrt
{2}})(|00\rangle-|11\rangle)_{bc}$, they possess the same bits,
which makes Bob  avoid detection.

Secondly, consider the intercept-resend in every even round (round
$2m$, $4\leq 2m\leq n$).

According to the BK protocol in Ref.\cite {BK}, after Alice's
operation $C_{a1}$, the whole state will be converted into
\begin{eqnarray}
(1/{\sqrt{2}})(|00\rangle_{a\overline{b}}|\overline{q}\rangle_{1,2}+
|11\rangle_{a\overline{b}}|\overline{1+q}\rangle_{1,2}) \otimes\nonumber\\
(1/{\sqrt{2}})(|00\rangle+|11\rangle)_{bc}\ (q_2=0),\nonumber\\
(1/{\sqrt{2}})(|01\rangle_{a\overline{b}}|\overline{q}\rangle_{1,2}+
|10\rangle_{a\overline{b}}|\overline{1+q}\rangle_{1,2}) \otimes\nonumber\\
(1/{\sqrt{2}})(|01\rangle+|10\rangle)_{bc}\ (q_2=1)
\end{eqnarray}

After receiving qubit \emph{1} and intercepting qubit \emph{2}, Bob
performs one single CNOT gate $C_{\overline{b} 1}$ on  qubits
$\overline{b}$ and  \emph{1} to produce the state
\begin{eqnarray}
(1/{\sqrt{2}})(|00\rangle+|11\rangle)_{a\overline{b}}\otimes
|\overline{q}\rangle_{1,2}\nonumber\\ \otimes
(1/{\sqrt{2}})(|00\rangle+
|11\rangle)_{bc}\ (q_2=0),\nonumber\\
(1/{\sqrt{2}})(|01\rangle+|10\rangle)_{a\overline{b}}\otimes
|\overline{1+q}\rangle_{1,2}\nonumber\\ \otimes
(1/{\sqrt{2}})(|01\rangle+ |10\rangle)_{bc}\ (q_2=1).
\end{eqnarray}
Then he  measures qubits \emph{1,2} in the Bell basis
\{$|\overline{0}\rangle_{12}$, $|\overline{1}\rangle_{12}$,
$(1/{\sqrt {2}})(|00\rangle-|11\rangle)_{12}$, $(1/{\sqrt
{2}})(|01\rangle-|10\rangle)_{12}$\}. He can gain one of two kinds
of measurement outcomes $|\overline{0}\rangle_{12}$ and
$|\overline{1}\rangle_{12}$, which denote bit 0 and  bit 1
 respectively. The
bit sent by Alice and that received by Bob are same when Alice and
Bob share the carriers  $(1/{\sqrt {2}})(|00\rangle +|11
\rangle)_{bc}$, but inverse when they share $(1/{\sqrt
{2}})(|01\rangle+|10\rangle)_{bc}$.

 After Bob has intercepted bit 0, he records $d_{2m}=0$, and then
records $e_{2m}=0,\psi=0$ or $e_{2m}=1,\psi=1$
 randomly, but after Bob has intercepted bit 1, he records $d_{2m}=1$, and then
records $e_{2m}=0,\psi=1$  or $e_{2m}=1,\psi=0$ randomly. The
message of bit $d_{2m}$ is split into bit $e_{2m}$ and bit $\psi$.
By the operation of resending bit, Bob sends bit $\psi$ to Charlie
with the help of the carriers $(1/{\sqrt
{2}}(|00\rangle+|11\rangle)_{bc}$ or $(1/ {\sqrt
{2}})(|01\rangle+|10\rangle)_{bc}$.

From the above analysis, we can see that when Alice, Bob, and
Charlie share the carriers  of $(1/{\sqrt
{2}})(|00\rangle+|11\rangle)_{a\overline {b}}\otimes (1/{\sqrt
{2}})(|00\rangle+|11\rangle)_{bc}$, Alice's bit $q_{2m}$ and Bob's
bit $d_{2m}$ are identical, and Bob's bit $\psi$ and Charlie's bit
are also identical. The sum of Alice's bit $q_{2m}$, Bob's bit
$e_{2m}$ and Charlie's bit is zero modulo 2, which makes Bob avoid
detection.

From the above analysis, we can also see that when Alice, Bob, and
Charlie share the carriers of $(1/{\sqrt
{2}})(|01\rangle+|10\rangle)_{a\overline {b}}\otimes (1/{\sqrt
{2}})(|01\rangle+|10\rangle)_{bc}$, Alice's bit $q_{2m}$ and Bob's
bit $d_{2m}$ are inverse, and Bob's bit $\psi$ and Charlie's bit are
also inverse. Two inverse operations  make the sum of Alice's bit
$q_{2m}$, Bob's bit $e_{2m}$ and Charlie's bit be zero modulo 2,
which makes Bob avoid detection.

It seems that the above cheat result cannot work entirely because
there are still two possibilities of Bob's eavesdropping bit string
\{$d_m,1\leq m \leq n$\}. However, according to the BK protocol in
Ref.\cite {BK}, the three legal parties have to compare a
subsequence of the data bits publicly to detect eavesdropping. Bob
announces the subsequence of the bit string \{$e_m,1\leq m\leq n$\}
publicly. Alice's bits or Charlie's bits will be leaked to Bob. More
specifically, as long as any even numbered data bit except the
second bit
 is announced, Bob can determine which kind of the two possible
carriers was  used and whether or not to flip the bits
$d_{2m}$($4\leq 2m\leq n$). If the public bits include the second
bit, only the second bit leads to error with probability 1/2, which
can be hidden in quantum noise. Otherwise, in term of the kind of
the carriers, Bob can know the bit $q_2$ precisely. By this means
Bob can obtain all the secret bits except for the little-probability
event that all the compared bits are odd numbered plus \emph{2}
numbered.

In conclusion, we have presented a cheat strategy that allows Bob
to gain all the secret bits before sharing, while introducing one
data bit error at most in the whole communication, which makes the
cheater avoid the detection by the communication parities.
Consequently this protocol cannot resist this type of attack.

This work is supported by the National Natural Science Foundation of
China, Grants No. 60373059; the National Laboratory for Modern
Communications Science Foundation of China; the National Research
Foundation for the Doctoral Program of Higher Education of China,
Grants No. 20040013007; the ISN Open Foundation; and  the Major
Research plan of the National Natural Science Foundation of
China(Grant No. 90604023).

\end{document}